\documentclass[envcountsame, runningheads]{llncs}

\usepackage[utf8]{inputenc}
\usepackage[T1]{fontenc}
\usepackage{lmodern}
\usepackage[hidelinks]{hyperref}
\usepackage{diagbox}

\usepackage[font=small,labelfont=bf]{caption}
\usepackage{subcaption}
\usepackage{amssymb}
\usepackage{amsmath}
\usepackage{stmaryrd}
\usepackage{tikz}
\usetikzlibrary{backgrounds, positioning, arrows, calc}
\usepackage{mathtools}
\usepackage[linesnumbered,vlined,noend, ruled]{algorithm2e}
\usepackage[inline]{enumitem}  
\usepackage[capitalise]{cleveref}
\usepackage{lineno}
\usepackage{booktabs}
\usepackage{multirow}
\usepackage{listings}
\usepackage{xcolor}
\usepackage[scaled=.85]{beramono}

\renewcommand{\cref}{\Cref}

\usepackage{todonotes}

\newcommand{\DD}{\ensuremath{\mathbb{D}}\xspace}

\newcommand{\QQ}{\ensuremath{\mathbb{Q}}\xspace}
\newcommand{\RR}{\ensuremath{\mathbb{R}}\xspace}

\newcommand{\A}{\mathcal{A}}
\newcommand{\B}{\mathcal{B}}
\newcommand{\C}{\mathcal{C}}

\newcommand{\trans}[3]{#1\xrightarrow[]{#2}#3}

\newcommand{\Entry}[1]{\begin{tabular}[c]{@{}c@{}} #1\end{tabular}}

\tikzstyle{state}=[thick,minimum size=18pt, circle,draw]
\tikzstyle{transition}=[->,thick,>=stealth,shorten >=1pt,shorten <=1pt]
\tikzstyle{final}=[after node path={ node[state, scale=.8] at (\tikzlastnode) {} }]
\tikzstyle{initial}=[after node path={
	[to path={[transition] (\tikztostart) -- (\tikztotarget)}]
	(\tikzlastnode)++(180:22pt) edge (\tikzlastnode)
}]
\tikzset{
	bg/.default={},
	bg/.style={execute at end picture={
			\begin{scope}[on background layer]
				\node[xshift=-1mm, yshift=-1mm] (sw) at (current bounding box.south west) {};
				\node[xshift=1mm, yshift=1mm] (ne) at (current bounding box.north east) {};
				\node[xshift=1mm, yshift=-1mm] (nw) at (current bounding box.north west) {};
				\fill[fill=black!10,rounded corners] (sw) rectangle (ne);
				
				\ifx&#1&\else
				\node[anchor=north east, xshift=2pt] at (nw) {#1};
				\fi
			\end{scope}
	}},
}

\newcommand{\Func}[1]{{\mathsf{#1}}}
\newcommand{\Val}{\Func{Val}}
\newcommand{\Inf}{\Func{Inf}}
\newcommand{\Sup}{\Func{Sup}}
\newcommand{\DSum}{\Func{DSum}}
\newcommand{\LimInf}{\Func{LimInf}}
\newcommand{\LimSup}{\Func{LimSup}}
\newcommand{\LimInfAvg}{\Func{LimInfAvg}}
\newcommand{\LimSupAvg}{\Func{LimSupAvg}}

\newcommand{\prefix}{\prec}

\newcommand{\suchthat}{\;\ifnum\currentgrouptype=16 \middle\fi|\;}
\let\st\suchthat

\newcommand{\safe}[1]{{\it SafetyCl}(#1)}

\newcommand{\CompClass}[1]{{\textsc{#1}}\xspace}
\newcommand{\PTime}{\CompClass{PTime}}

\newcommand{\PSpaceC}{\CompClass{PSpace}-complete\xspace}
\newcommand{\PSpaceH}{\CompClass{PSpace}-hard\xspace}
\newcommand{\ExpSpace}{\CompClass{ExpSpace}}

\title{Automating the Analysis of Quantitative Automata with QuAK\thanks{This work was supported in part by the ERC-2020-AdG 101020093.}} 
\titlerunning{Automating the Analysis of Quantitative Automata with QuAK}

\author{
	Marek~Chalupa\inst{1}\orcidID{0000-0003-1132-5516} \and 
	Thomas~A.~Henzinger\inst{1}\orcidID{0000-0002-2985-7724} \and 
	Nicolas~Mazzocchi\inst{2}\orcidID{0000-0001-6425-5369} \and 
	N.~Ege~Sara\c{c}\inst{1}\orcidID{0009-0000-2866-8078}
}
\authorrunning{M.~Chalupa \and T.~A.~Henzinger \and N.~Mazzocchi \and N.~E.~Sara\c{c}} %

\institute{
	Institute of Science and Technology Austria (ISTA), Austria\\
	\email{\{mchalupa,tah,esarac\}@ista.ac.at}
	\and
	Slovak University of Technology in Bratislava, Slovak Republic\\
	\email{nicolas.mazzocchi@stuba.sk}
}

\date{}

\makeatletter 
\def\orcidID#1{\smash{\href{http://orcid.org/#1}{\protect\raisebox{-1.25pt}{\protect\includegraphics{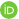}}}}}    
\makeatother

\begin{document}
	\maketitle
	
	\begin{abstract}
		Quantitative automata model beyond-boolean aspects of systems: every execution is mapped to a real number by incorporating weighted transitions and value functions that generalize acceptance conditions of boolean $\omega$-automata.
		Despite the theoretical advances in systems analysis through quantitative automata, the first comprehensive software tool for quantitative automata (Quantitative Automata Kit, or QuAK) was developed only recently.
		QuAK implements algorithms for solving standard decision problems, e.g., emptiness and universality, as well as constructions for safety and liveness of quantitative automata.
		We present the architecture of QuAK, which reflects that all of these problems reduce to either checking inclusion between two quantitative automata or computing the highest value achievable by an automaton---its so-called top value.
		We improve QuAK by extending these two algorithms with an option to return, alongside their results, an ultimately periodic word witnessing the algorithm's output, as well as implementing a new safety-liveness decomposition algorithm that can handle nondeterministic automata, making QuAK more informative and capable.
	\end{abstract}
	
	\section{Introduction}
	
	Traditional system behavior analysis categorizes system behaviors as correct or incorrect.
	However, modern systems require more nuanced approaches to address performance and robustness criteria.
	Quantitative automata generalize boolean $\omega$-automata by adding rational-valued weights to their transitions and using value functions (instead of acceptance conditions) that accumulate infinite weight sequences into single values.
	Common value functions include $\Inf$, $\Sup$, $\LimInf$, and $\LimSup$ (respectively generalizing safety, reachability, co-Büchi and Büchi acceptance conditions), as well as $\DSum$ (discounted sum), $\LimInfAvg$, and $\LimSupAvg$ (limit average a.k.a. mean payoff).
		
	Decision problems for boolean automata extend naturally to quantitative automata.
	For example, a quantitative automaton $\A$ is nonempty with respect to a rational number $v$ iff $\A$ maps some infinite word $w$ to a value at least $v$~\cite{DBLP:journals/tocl/ChatterjeeDH10}.
	These problems are closely related to game theory~\cite{DBLP:journals/corr/abs-2305-10546} and enable reasoning about quantitative system aspects.
	Although quantitative automata have been extensively studied from a theoretical perspective~\cite{DBLP:conf/csl/DegorreDGRT10,DBLP:journals/corr/BokerH14,DBLP:conf/lics/BokerHO15,DBLP:conf/concur/MichaliszynO19,DBLP:conf/ijcai/MichaliszynO21,DBLP:conf/lics/Boker24,DBLP:conf/lics/HenzingerS21,DBLP:conf/rv/HenzingerMS22,DBLP:conf/fossacs/HenzingerMS23,DBLP:conf/concur/BokerHMS23} and these works could significantly impact their practical verification, until recently, there was no general software tool for their analysis.
	
	Quantitative Automata Kit (QuAK)~\cite{chalupa2024quakquantitativeautomatakit} is the first general tool for quantitative automata analysis.
	It currently supports several automaton types ($\Inf$, $\Sup$, $\LimInf$, $\LimSup$, $\LimInfAvg$, $\LimSupAvg$) and provides decision procedures for fundamental problems such as emptiness, universality, inclusion, and safety.
	
	We present an improved version of QuAK: (i) the safety-liveness decompositions are extended to handle nondeterministic automata, and (ii) the inclusion and top value algorithms are extended with capabilities to return a witness---an ultimately periodic word explaining the algorithm's output.
	For checking inclusion, i.e., whether $\A(w) \leq \B(w)$ for all words $w$, the witness $\hat{w}$ is a word such that $\A(\hat{w}) > \B(\hat{w})$.
	For computing top value, i.e., $\top_{\A} = \sup_{w \in \Sigma^\omega} \A(w)$, the witness $\widehat{w}$ is a word such that $\A(\hat{w}) = \top_{\A}$.
	Since all the other procedures are reduced to either inclusion checking or top value computation, these extensions significantly improve QuAK's informativeness for analyzing quantitative automata.

	Several approaches extend system modeling beyond boolean aspects.
	One uses multi-valued truth domains~\cite{DBLP:conf/cav/BrunsG99,DBLP:conf/cav/ChechikGD02}, while another relies on weighted automata~\cite{DBLP:journals/iandc/Schutzenberger61b}, where numerical weights are assigned to transitions and accumulated via semiring operations.
	Tools such as Vaucanson~\cite{DBLP:conf/wia/LombardyPRS03}, Vcsn~\cite{DBLP:conf/wia/DemailleDLS13}, and Awali~\cite{Awali2.3} support weighted automata analysis.
	Other approaches address digital-analog interactions~\cite{DBLP:journals/iandc/AlurH93,DBLP:journals/tcs/AlurD94,DBLP:conf/lics/Henzinger96}, with tools like UPPAAL~\cite{DBLP:journals/sttt/LarsenPY97} and HyTech~\cite{DBLP:conf/hybrid/HenzingerH94a}.
	Signal temporal logic~\cite{DBLP:conf/formats/MalerN04} allows reasoning about specification satisfaction degrees, as implemented in Breach~\cite{DBLP:conf/cav/Donze10}, S-TaLiRo~\cite{DBLP:conf/tacas/AnnpureddyLFS11}, and RTAMT~\cite{DBLP:journals/sttt/YamaguchiHN24}.
	Probabilistic verification handles uncertainties, as implemented in PRISM~\cite{DBLP:conf/cpe/KwiatkowskaNP02} and STORM~\cite{DBLP:conf/cav/DehnertJK017}.
	
	\section{Quantitative Automata} \label{sec:definitions}
	
	Let $\Sigma$ be a finite alphabet of letters.
	We denote by $\Sigma^*$ (resp. $\Sigma^\omega$) the set of all finite (resp. infinite) words over $\Sigma$.
	For $w \in \Sigma^\omega$ and $u \in \Sigma^*$, we write $u \prec w$ when $u$ is a prefix of $w$.
	A \emph{value domain} $\DD$ is a nontrivial complete lattice.
	A \emph{quantitative property} is a total function $\varPhi : \Sigma^\omega \to \DD$.
		
	We study quantitative automata, which define a subset of quantitative properties on totally-ordered value domains.
	Formally, a \emph{nondeterministic quantitative automaton} (or simply an \emph{automaton}) is a tuple $\A = (\Sigma,Q,s,\delta)$ where $\Sigma$ is a finite alphabet, $Q$ is a finite nonempty set of states, $s \in Q$ is the initial state, and $\delta \colon Q \times \Sigma \to 2^{\QQ \times Q}$ is a finite transition function over weight-state pairs~\cite{DBLP:journals/tocl/ChatterjeeDH10}.
	A \emph{transition} is a tuple $(q, \sigma, x, q') \in Q \times \Sigma \times \QQ \times Q$ such that $(x,q') \in \delta(q,\sigma)$, denoted $\trans{q}{\sigma:x}{q'}$.
	The weight of a transition $t = (q, \sigma, x, q')$ is denoted by $\gamma(t) = x$.
	An automaton $\A$ is \emph{deterministic} when $|\delta(q,a)| = 1$ for every $q \in Q$ and $a \in \Sigma$, and it is \emph{total} (a.k.a. \emph{complete}) when $|\delta(q,a)| \geq 1$ for every $q \in Q$ and $a \in \Sigma$.
	We require quantitative automata to be total.

	A run of $\A$ on an infinite word $w = a_0 a_1 \ldots$ is an infinite sequence $\rho = \trans{q_0}{a_0:x_0}{q_1}\trans{}{a_1:x_1}{q_2}\ldots$ of transitions where $q_0 = s$ and $(x_i, q_{i+1}) \in \delta(q_i, a_i)$ for each integer $i \geq 0$.
	Since each transition has a weight, a run $\rho = t_0 t_1 \ldots$ produces an infinite sequence $\gamma(\rho) = \gamma(t_0) \gamma(t_1) \ldots$ of weights.
	A value function $\Val : \QQ^\omega \to \RR$ maps infinite weight sequences to real values.
	A $\Val$-automaton is a quantitative automaton equipped with the value function $\Val$, i.e., where each run $\rho$ is mapped to the value obtained by applying $\Val$ to its weight sequence $\gamma(\rho)$.
	Given a $\Val$-automaton $\A$, the value of an infinite word $w$ is $\A(w) = \sup\{\Val(\gamma(\rho)) \st \text{$\rho$ is a run of $\A$ on $w$}\}$.
	The \emph{top value} of an automaton $\A$ is $\top_{\A} = \sup_{w \in \Sigma^\omega} \A(w)$, and its \emph{bottom value} is $\bot_{\A} = \inf_{w \in \Sigma^\omega} \A(w)$.
	We consider the below value functions over an infinite sequence $x = x_0 x_1 \ldots$ of rational weights.

	\begin{multicols}{2}
		\begin{itemize}
			\item $\displaystyle\Inf(x) = \inf\limits_{n \geq 0} x_n$
			\item $\displaystyle\LimInf(x) = \liminf\limits_{n\to\infty} x_n$
			\item $\displaystyle\LimInfAvg(x) = \liminf\limits_{n\to\infty} \left(\frac{1}{n} \sum_{i=0}^{n-1} x_i\right)$
			\item $\displaystyle\Sup(x) = \sup\limits_{n \geq 0} x_n$
			\item $\displaystyle\LimSup(x) = \limsup\limits_{n\to\infty} x_n$
			\item $\displaystyle\LimSupAvg(x) = \limsup\limits_{n\to\infty} \left(\frac{1}{n} \sum_{i=0}^{n-1} x_i\right)$
		\end{itemize}
	\end{multicols}
	
	\begin{itemize}
		\item For a discount factor $\displaystyle\lambda\in\QQ\cap(0,1)$, $\DSum_{\lambda}(x) = \sum_{i\geq 0} \lambda^i  x_i$
	\end{itemize}

\begin{figure}[t]
		\begin{minipage}{.49\linewidth}
			\scalebox{.7}{
				\begin{tikzpicture}[bg={($\A$)}, node distance=1.75cm]
					\node[state,label=center:$q_0$] (q0) {};
					\draw[transition](q0.90)++(90:13pt) -- (q0);
					\node[state, left of = q0, label=center:$q_1$] (q1) {};
					\node[state, right of = q0, label=center:$q_2$] (q2) {};
					\node[state, right of = q2, label=center:$q_3$] (q3) {};
					
					\path[transition]
					(q0) edge node[above] {$\textit{hi}:6$} (q1)
					(q0) edge node[above] {$\textit{lo}:3$} (q2)
					(q1) edge[loop above] node[above] {$\Sigma:6$} (q1)
					(q2) edge[loop above] node[above] {$\textit{lo}:2$} (q2)
					(q2) edge node[below] {$\textit{hi}:5$} (q3)
					(q3) edge[bend right = 20] node[above] {$\textit{lo}:3$} (q2)
					(q3) edge[loop above] node[above] {$\Sigma:4$} (q3)
					;
				\end{tikzpicture}
			}
			\bigskip\\
			\scalebox{.7}{
				\begin{tikzpicture}[bg={($\B$)}, node distance=1.75cm]
					\node[state,label=center:$q_0$] (q0) {};
					\draw[transition](q0.90)++(90:13pt) -- (q0);
					\node[state, left of = q0, label=center:$q_1$] (q1) {};
					\node[state, right of = q0, label=center:$q$] (q) {};
					
					\path[transition]
					(q0) edge node[above] {$\textit{hi}:6$} (q1)
					(q0) edge node[above] {$\textit{lo}:4$} (q2)
					(q1) edge[loop above] node[above] {$\Sigma:6$} (q1)
					(q) edge[loop above] node[above] {$\Sigma:4$} (q)
					;
				\end{tikzpicture}
			}
		\end{minipage}
		\begin{minipage}{.49\linewidth}\centering
			\scalebox{.7}{
				\begin{tikzpicture}[bg={($\C$)}, node distance=1.75cm]
					\node[state,label=center:$q_0$] (q0) {};
					\draw[transition](q0.90)++(90:13pt) -- (q0);
					\node[state, left of = q0, label=center:$q_1$] (q1) {};
					\node[state, right of = q0, label=center:$q_2$] (q2) {};
					\node[state, right of = q2, label=center:$q_3$] (q3) {};
					
					\path[transition]
					(q0) edge node[above] {$\textit{hi}:6$} (q1)
					(q0) edge node[above] {$\textit{lo}:3$} (q2)
					(q1) edge[loop above] node[above] {$\Sigma:6$} (q1)
					(q2) edge[loop above] node[above] {$\textit{lo}:2$} (q2)
					(q2) edge node[below] {$\textit{hi}:5$} (q3)
					(q3) edge[bend right = 20] node[above] {$\textit{lo}:3$} (q2)
					(q3) edge[loop above] node[above] {$\Sigma:4$} (q3)
					;

					\node[state, yshift=-1.5cm, label=center:$q'_1$] (qq1) at (q1) {};
					\node[state, yshift=-1.5cm, label=center:$q'_3$] (qq3)at (q3) {};
					\node[state, yshift=-3cm, xshift=0.75cm, label=center:$q_\bot$] (qbot) at (q0){};
					
					\path[transition]
					(q1) edge  node[right] {$\textit{hi}:6$} (qq1)
					(qq1) edge[loop below] node[below] {$\textit{hi}:6$} (qq1)
					(q3) edge[pos=.18]  node[below left] {$\textit{hi}:6$} (qq3)
					(qq3) edge[loop below] node[below] {$\textit{hi}:6$} (qq3)
					(qq1) edge[bend right=20] node[above right] {$\Sigma:2$} (qbot)
					(qq3) edge[bend left=20] node[above left] {$\Sigma:2$} (qbot)
					(qbot) edge[loop above] node[above] {$\Sigma:2$} (qbot)
					;	
				\end{tikzpicture}
			}
		\end{minipage}
		\caption{
			A nondeterministic $\LimInfAvg$-automaton $\A$ over the alphabet $\Sigma = \{\textit{hi}, \textit{lo}\}$, modeling the power consumption of a device where starting with high-power mode is not reversible, its safety closure $\B$, and its liveness component $\C$ in a corresponding decomposition~\cite{boker2024safetylivenessquantitativeproperties}.
		}
		\label{fig:avgDecomp}
	\end{figure}
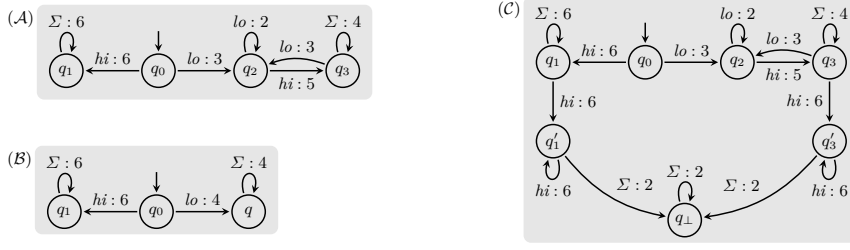

	The automaton $\A$ in \Cref{fig:avgDecomp} shows a $\LimInfAvg$ automaton modeling the long-term average power consumption of a device with two operating modes.
	
	\subsubsection{Quantitative Automata Problems} \label{sec:problems}
	We describe the standard decision problems of quantitative automata as well as the problems related to their safety and liveness.
	The complexity results are summarized in \Cref{tbl:complexity}.

	An automaton $\A$ is \emph{nonempty} (resp. \emph{universal}) with respect to a threshold $v \in \QQ$ iff $\A(w) \geq v$ for some (resp. all) $w \in \Sigma^\omega$.
	Nonemptiness (resp. universality) is closely related to computing an automaton's top value (resp. bottom value): $\A$ is nonempty (resp. universal) with respect to $v \in \QQ$ iff $\top_{\A} \geq v$ (resp. $\bot_{\A} \geq v$).
	An automaton $\A$ is \emph{included in} (resp. \emph{equivalent to}) an automaton $\B$ iff $\A(w) \leq \B(w)$ (resp. $\A(w) = \B(w)$) for all $w \in \Sigma^\omega$.
	An automaton $\A$ is \emph{constant} iff there exists $c \in \RR$ such that $\A(w) = c$ for all $w \in \Sigma^\omega$.
	This problem is closely related to safety and liveness of quantitative automata, as we discuss below.
		
		\begin{table}[t]\centering
		\bgroup \renewcommand*{\arraystretch}{1.2}
		\footnotesize \scalebox{0.75}{\begin{tabular}{|c||c|c|c|c|}
				\hline
				\phantom{\textbf{(b)}}& $\Inf$ & $\Sup$, $\LimInf, \LimSup$ & $\LimInfAvg, \LimSupAvg$ & $\DSum$ \\
				\hline\hline
				Nonemptiness check &
				\multicolumn{4}{c|}{\PTime}\\
				\hline
				Universality check & \multicolumn{2}{c|}{\PSpaceC} & Undecidable & Open\\
				\hline
				Inclusion check & \multicolumn{2}{c|}{\PSpaceC} & Undecidable & Open\\
				\hline
				Equivalence check & \multicolumn{2}{c|}{\PSpaceC} & Undecidable & Open\\		
				\hline
				\Entry{Top value\vspace{-0.5em}\\computation}&
				\multicolumn{4}{c|}{\PTime}\\
				\hline
				\Entry{Safety closure\vspace{-0.5em}\\construction} &
				$O(1)$ & \multicolumn{2}{c|}{\PTime}& $O(1)$ \\
				\hline
				\Entry{Safety-liveness\vspace{-0.5em}\\decomposition}& 
				$O(1)$ & \multicolumn{2}{c|}{\PTime}
				& $O(1)$ \\
				\hline
				Safety check& 
				$O(1)$ & \PSpaceC & \ExpSpace; \PSpaceH & $O(1)$ \\
				\hline
				Liveness check& \multicolumn{4}{c|}{\PSpaceC}\\
				\hline
				\Entry{Constant-function\vspace{-0.5em}\\check}&
				\multicolumn{4}{c|}{\PSpaceC}\\
				\hline
		\end{tabular}}
		\egroup
		\vspace{1em}
		\caption{
			The complexity of performing the operations on the left column with respect to nondeterministic automata with the value function on the top.
			The decidability results in the top five rows are shown in~\cite{DBLP:conf/vmcai/KupfermanL07,DBLP:journals/tocl/ChatterjeeDH10} and undecidability in~\cite{DBLP:conf/csl/DegorreDGRT10,DBLP:conf/concur/ChatterjeeDEHR10,DBLP:journals/tcs/HunterPPR18}.
			All the results in the bottom five rows are shown in~\cite{boker2024safetylivenessquantitativeproperties}.
			All the operations are computable in \PTime for deterministic automata.
			\vspace{-2em}
		}
		\label{tbl:complexity}
	\end{table}
	
	Quantitative safety generalizes the boolean view by considering membership hypotheses in the form of lower bound queries: a property is safe iff every wrong membership hypothesis has a finite witness for the violation.
	Formally, a quantitative property $\varPhi : \Sigma^\omega \to \DD$ is \emph{safe} iff for every $w \in \Sigma^\omega$ and $v \in \DD$ with $\varPhi(w) \not \geq v$, there exists a finite prefix $u \prefix w$ such that $\sup_{w' \in \Sigma^\omega} \varPhi(u w') \not \geq v$~\cite{DBLP:conf/fossacs/HenzingerMS23}.
	Moreover, an automaton $\A$ is safe iff the quantitative property defined by $\A$ is safe.
	Given a quantitative property $\varPhi : \Sigma^\omega \to \DD$, its \emph{safety closure} is defined as $\safe{\varPhi}(w) = \inf_{u \prefix w} \sup_{w' \in \Sigma^\omega} \varPhi(u w')$ and is the least safety property that bounds $\varPhi$ from above~\cite{DBLP:conf/fossacs/HenzingerMS23}.
	As expected, a property $\varPhi$ is safe iff $\varPhi(w) = \safe{\varPhi}(w)$ for all $w \in \Sigma^\omega$, and we can compute the safety closure of an automaton $\A$---the automaton $\safe{\A}$ that expresses the safety closure of the property defined by $\A$.
	While this characterization  is useful for some classes of quantitative automata, the equivalence problem is undecidable for $\LimInfAvg$ and $\LimSupAvg$ automata.
	For these, the safety problem is still decidable by a reduction to their constant-function problem~\cite{DBLP:conf/concur/BokerHMS23}.
	
	Quantitative liveness extends the membership-based view: a quantitative property $\varPhi : \Sigma^\omega \to \DD$  is live iff for every word (whose value is less than $\top = \sup \DD$) there exists a wrong membership hypothesis without a finite witness for the violation.
	Formally, a quantitative property $\varPhi : \Sigma^\omega \to \DD$ is \emph{live} iff for all $w \in \Sigma^\omega$, if $\varPhi(w) < \top$, then there exists a value $v \in \DD$ such that $\varPhi(w) \not \geq v$ and for all prefixes $u \prec w$, we have $\sup_{w' \in \Sigma^\omega} \varPhi(uw') \geq v$~\cite{DBLP:conf/fossacs/HenzingerMS23}.
	Moreover, an automaton $\A$ is live iff the quantitative property defined by $\A$ is live. 
	For the common classes of quantitative automata, deciding liveness reduces to the constant-function problem: an automaton $\A$ is live iff $\safe{\A}$ is constant~\cite{DBLP:conf/concur/BokerHMS23}.
	Just like every boolean property is the intersection of its safety closure and a liveness property, every quantitative property is the pointwise minimum of its safety closure and a liveness property~\cite{DBLP:conf/fossacs/HenzingerMS23}.
	Recently, it was proved that all the common classes of automata can be decomposed into its safety closure and a liveness property~\cite{boker2024safetylivenessquantitativeproperties}.
	Consider the  automaton $\A$, its safety closure $\B$, and its liveness part $\C$ as defined in \cref{fig:avgDecomp}.
	In $\B$, each strongly connected component (SCC) of $\A$ is assigned the highest value achievable within the component, representing the greatest among the lower bound hypotheses that cannot be refuted by any finite prefix.
	The liveness part $\C$ consists of three components: the upper part is a copy of $\A$ (ensuring $\C$ can have runs with the same value as $\A$); the middle part contains a $\top_{\A}$-weighted copy of the highest-valued cycle in each SCC (enabling $\C$ to achieve high-valued runs when $\A$ and $\B$ agree); and the lower part includes a sink state looping with the lowest weight of $\A$ (allowing $\C$ to ``escape'' the middle part and realize a value using the upper part).
	
	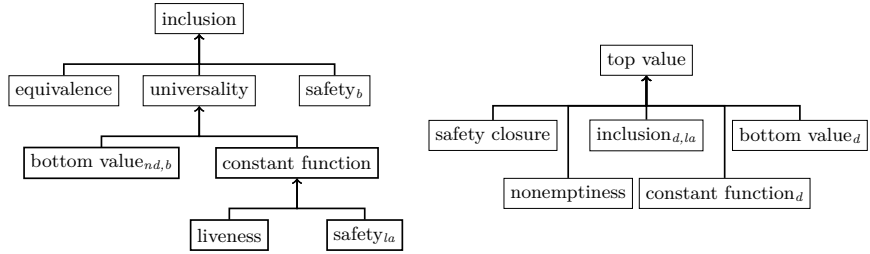
\begin{figure}[t]
	\centering
	\noindent
	\begin{subfigure}[c]{.40\linewidth}
		\scalebox{0.8}{
\begin{tikzpicture}[
	every node/.style={draw, rectangle, minimum width=1cm, minimum height=0.5cm},
	level/.style={sibling distance=2.25cm, level distance=1.2cm},
	edge from parent/.style={draw, <-, thick},
	edge from parent path={(\tikzparentnode.south) -- ++(0,-0.5cm) -| (\tikzchildnode.north)} 
	]
	\node {inclusion}
	child {
		node {equivalence}
	}
	child {
		node {universality}
		child {
			node[align=center, xshift=-0.5cm] {bottom value$_{\textit{nd,b}}$}
		}
		child {
			node[align=center, xshift=0.5cm] {constant function}
			child {
				node {liveness}
			}
			child {
				node {safety$_{\textit{la}}$}
			}
		}
	}
	child {
		node {safety$_{\textit{b}}$}
	};
\end{tikzpicture}

		}
	\end{subfigure}
	\noindent
	\hspace{1.5em}
	\begin{subfigure}[c]{.5\linewidth}
		\scalebox{0.8}{
		\begin{tikzpicture}[
			every node/.style={draw, rectangle, minimum width=1cm, minimum height=0.5cm},
			level distance=1.25cm,
			sibling distance=2.55cm,
			edge from parent/.style={draw, <-, thick},
			edge from parent path={(\tikzparentnode.south) -- ++(0,-0.5cm) -| (\tikzchildnode.north)} 
			]
			\node[align=center](top) {top value}
			child {
				node[align=center] {safety closure}
			}
			child {
				node {inclusion$_{\textit{d,la}}$}
			}
			child {
				node[align=center] {bottom value$_{\textit{d}}$}
			};
			
			\node[below=1.7cm of top, xshift=-1.3cm] (nonempt) {nonemptiness};
			\draw[<-, thick] (top.south) -- ++(0,-0.5cm) -| (nonempt.north);
			
			\node[below=1.7cm of top, xshift=1.3cm, align=center] (cfd) {constant function$_{\textit{d}}$};
			\draw[<-, thick] (top.south) -- ++(0,-0.5cm) -| (cfd.north);
		\end{tikzpicture}
		}
	\end{subfigure}%
	\small
	\caption{
		Reductions of quantitative automata problems in QuAK.
		The subscript \emph{b} stands for basic (i.e., $\Val \in \{\Inf, \Sup, \LimInf, \LimSup\}$), \emph{la} for limit-average (i.e., $\Val \in \{\LimInfAvg, \LimSupAvg\}$), \emph{d} for deterministic, and \emph{nd} for nondeterministic.
		For example, checking safety of limit-average automata (safety$_\textit{la}$) reduces to their constant-function problem, which reduces to universality of $\LimInf$ automata.
	}
	\label{fig:reductions}
\end{figure}
		
	\section{QuAK Overview and Usage} \label{sec:overview}
	
	QuAK is written in C++ using the standard library, and the source code is available online~\cite{quak}.
	It can be used both as a C++ library and a stand-alone tool through the command-line interface -- see our project repository for instructions.
	For $\Inf$, $\Sup$, $\LimInf$, $\LimSup$, $\LimInfAvg$, and $\LimSupAvg$ automata, QuAK implements the operations listed in \Cref{tbl:complexity} and a monitoring procedure where the monitor maintains the maximal and minimal possible values or a running average.
	The problems handled in QuAK reduce to either the inclusion problem or the top value computation, as shown in \Cref{fig:reductions}.
	For the details of these reductions, we refer the reader to~\cite{DBLP:journals/tocl/ChatterjeeDH10,boker2024safetylivenessquantitativeproperties,chalupa2024quakquantitativeautomatakit}.
	
	We improve over QuAK's initial version~\cite{chalupa2024quakquantitativeautomatakit} in two ways.
	First, a safety-liveness decomposition for nondeterministic automata is implemented, following a new result that provides a decomposition for nondeterministic automata with prefix-independent value functions, namely, $\LimInf$, $\LimSup$, $\LimInfAvg$, and $\LimSupAvg$ automata~\cite[Thm.~9.11]{boker2024safetylivenessquantitativeproperties}.
	Second, the inclusion-checking and the top-value computation algorithms are extended with an option to return a witness (for negative instance of inclusion and all instances of top value).
	Since all other problems reduce to these two (see \cref{fig:reductions}), the ability to generate witnesses makes QuAK more informative for analyzing quantitative automata.
	Inclusion checking, a central component of QuAK, is implemented using an antichain-based algorithm, extending FORKLIFT for Büchi automata~\cite{DBLP:conf/cav/DoveriGM22}, with details and performance benefits discussed in~\cite{chalupa2024quakquantitativeautomatakit}.
	As this algorithm systematically searches for counterexamples, it inherently supports witness construction for negative instances.
	Top value computation is based on graph-theoretic algorithms~\cite{DBLP:journals/tocl/ChatterjeeDH10}, with witness generation achieved via backtracking pointers.
	These improvements, along with other features of QuAK, were validated through unit testing, random testing, and cross-validation with existing implementations.

	QuAK reads and constructs automata from text files.
	Each automaton is represented as a list of transitions of the format \texttt{a : v, q -> p} which encodes a transition from state $q$ to state $p$ with letter $a$ and weight $v$.
	The initial state of the input automaton is the source state of the first transition in its text file.	

	Recall the nondeterministic limit-average automaton $\A$ and its safety-liveness decomposition from \Cref{fig:avgDecomp}.
	The first three lines of the code snippet in \Cref{fig:program} construct the automata $\A$, $\B$, and $\C$ as presented in \Cref{fig:avgDecomp}.
	The nonemptiness check returns \texttt{false}, and \texttt{witEmpt} points to an array storing $u = \textit{hi}$ and $v = \textit{hi}$ as $\top_{\A} = \A(\textit{hi}^\omega) = 6$.
	Similarly, the safety check returns \texttt{false} and \texttt{witSafe} points to an array storing $u = v = \textit{lo}$ as $\B(\textit{lo}^\omega) = 4$ and $\A(\textit{lo}^\omega) = 2$.

	\begin{figure}[t]
		\centering
		\begin{lstlisting}[gobble=1]
      Word* witEmpt, witSafe;
      Automaton* A = new Automaton("A.txt");
      Automaton* B = safetyClosure(A, LimInfAvg);
      Automaton* C = livenessComponent(A, LimInfAvg);
      bool flagEmpt = A->isNonEmpty(LimInfAvg, 5, &witEmpt);
      bool flagSafe = A->isSafe(LimInfAvg, &witSafe);
		\end{lstlisting}
		\caption{
			An example usage of QuAK as a C++ library.
			The functions \texttt{isNonEmpty} and \texttt{isSafe} now take an additional (optional) parameter for storing the stem and the period of the ultimately periodic word witnessing the algorithms' outputs.
			}
		\label{fig:program}
	\end{figure}

	\section{Conclusion}
	We presented an improved version of QuAK, our software tool for automating quantitative automata analysis, which extends the functionality introduced in~\cite{chalupa2024quakquantitativeautomatakit}.
	Future work aims to improve the tool's scalability and applicability while exploring more efficient verification methods.
	One promising avenue is the development of symbolic approaches to efficiently manage large state spaces.
	Another key direction involves extending the tool to support additional formalisms, such as various types of discounted-sum automata~\cite{DBLP:conf/lics/Boker24}, mean-payoff automaton expressions~\cite{DBLP:conf/concur/ChatterjeeDEHR10}, and nested quantitative automata~\cite{DBLP:journals/tocl/ChatterjeeHO17}.
	In parallel with these efforts, developing novel verification methods specifically tailored to the safety fragments of expressive quantitative formalisms presents an exciting research direction.
	
	\bibliographystyle{splncs04}
	\bibliography{main}
\end{document}